\documentstyle[preprint,prb,aps]{revtex}
\newcommand{\f}{$f$}
\newcommand{\dd}{{\rm d}}
\newcommand{\kk}{{\bf k}}
\newcommand{\rr}{{\bf r}}
\newcommand{\RR}{{\bf R}}
\newcommand{\qq}{{\bf q}}

\newcommand{\HH}{{\cal H}}
\begin{document}
\draft
\date{\today}
\title{Poisson equation and self-consistent periodical Anderson model}

\author{U. Lundin$^1$, I. Sandalov$^{1,2}$, O. Eriksson$^1$}
\address{$^1$Condensed Matter Theory group, Uppsala University, Box 530,
SE-751 21 Uppsala, Sweden\\
$^2$Dept.\ of Physics, Link\"oping University,
SE-581 83 Link\"{o}ping, Sweden}

\maketitle

\begin{abstract}
We show that the formally exact expression for the free energy (with a 
non-relativistic Hamiltonian) for the correlated metal generates the
Poisson equation within the
saddle-point approximation for the electric potential, where 
the charge density automatically includes correlations. 
In this approximation the problem is reduced to the self-consistent
periodical Anderson model (SCPAM). The parameter of the mixing
interaction in this formulation have to be found self-consistently
together
with the correlated charge density. The factors, calculated by 
Irkhin, for the mixing
interaction, which reflect the structure of the many-electron states of
the \f-ion involved, arise automatically in this formulation 
and are quite sensitive to the specific element we are interested in.
We also discuss the definitions of
the mixing interaction for the mapping from {\it ab initio} to model
calculations. 
\end{abstract}

\maketitle

\section{Introduction}
The local density approximation (LDA) to
density functional theory (DFT) is working surprisingly well in many
cases for which is is expected not to work at all. One of its most
important 
features is that the self-consistent procedure provides a quite accurate
distribution of the charge density, calculated from the Kohn-Sham 
equation~\cite{kohn-sham} which is more
accurate than the Poisson equation 
It is also important to understand why
the form of the potential which has been derived from the theory of a
\emph{homogeneous} electron gas with the charge density which includes
contributions from the localized electrons (the extreme case of
non-homogeneity) works so well. Below we will show that, at least, the 
Poisson
equation can be formulated also in the case of strongly correlated
system,
where some part of the electrons are either fully or partly localized. 
One the one hand, physically it
is clear that the role played by the localized (core)
electrons in the formation of the potential is the screening of
the nuclear 
potential experienced by the conduction electrons (which contribute to
the cohesive energy). The 
localized electrons do not contribute to the cohesive energy and,
therefore, the error, coming from an 
insufficient description of the core electrons 
is not so essential for
the properties derived from a calculation of the total energy at zero
temperature. On the other hand, the experience accumulated using the
LDA-DFT
shows that it fails to describe the properties, which require 
information about (quasi)localized electrons, like photo-electron
spectroscopy experiments, or exchange interaction between localized
electrons in magnetic insulators, semiconductors, etc. In these cases, 
methods either completely based on the field theory or the ones 
combining the field-theory and DFT, are desirable. The models 
often used are the Anderson impurity and periodical models and here we
will discuss mainly the periodical model. 
There are two ways for providing methods combining DFT and field theory. 
The first way is the following. The field operators can be constructed 
using the 
functions generated within a LDA-DFT calculation and 
a correction, constructed from the difference 
$\HH^{int}=\HH^{full}-\HH^{LDA}$, can be used to correct the 
electronic structure 
generated by the initial LDA calculation~\cite{DT}. 
The motivation 
for using the LDA-DFT basis here is that one can expect compensation by the
LDA potential from those part of the self-energy for the conduction 
electrons, which
corresponds to a static random phase approximation~\cite{kotani}. 
This way, however, 
involves complex calculations of the matrix elements of the Coulomb 
interaction and frequency-dependent magnitudes, like the self-energy,
that in practice such calculations are quite hard to
perform~\cite{steiner,manghi}. 
The other way, 
much more often used, is some mapping to the Anderson or Hubbard models.
Then, two difficulties arise. {\it First}, the question about double 
counting 
of some of the interactions, and also how to calculate the parameters of
the 
model which is chosen for treating correlations. 
{\it Second}, the 
model calculations often involve a redistribution of spectral weights 
between low- and high-energy regions and a redistribution of the 
charge density caused by it. The latter is never taken into account in the
model calculations. 
It is especially important since any redistribution of charge 
involves a large Coulomb energy. This is especially important when
the 
Anderson model is used for discussing magnetic properties. These
properties are 
determined by the effective exchange integral $\sim V^{2}/U$ (where $V$
is the 
mixing and $U$ is the Hubbard on-site repulsion) and, therefore,
involves 
\emph{small} energies. Calculations of this small energy difference, 
having 
neglected a possibly greater Coulomb energy, can easily lead to a 
misinterpretation of the experimental data. It is also important that 
the 
mixing interaction is representation dependent and, therefore, for the 
description of a real system within a model it is important to define 
clearly what is mixing interaction for that special case. 
This shows the need for a formulation 
which allows for a self-consistent calculation of the parameters 
of the model together with the charge density. Such an opportunity 
arise in 
a natural way if one starts with the full Hamiltonian and treat the 
single-ion Coulomb interactions in some approximation which takes 
into account the strong local electron correlations. Such a scheme has 
been 
suggested earlier~\cite{sand_fil}, however, the strong electron 
correlations 
(SEC) have been treated within the slave-boson technique which at
present
seems to be unsatisfactory for reasons which we will not discuss here
(see, Ref.~\cite{sand_richt}). Here we will discuss the parameters of
the
Anderson model within the same scheme, using the definitions of
the operators in a non-orthogonal basis set
different
from Ref.~\cite{sand_fil}, but coinciding with the ones used in 
Ref.~\cite{DT} and~\cite{II}. We have discussed a way
to calculate the single-site Coulomb repulsion parameter, Hubbard
$U$, 
earlier~\cite{DT,II,urban_model}. Here we pay attention mainly
to the mixing interaction. The organization of the paper is as follows.
In section II we rewrite the Hamiltonian in an non-orthogonal basis set, 
and construct the many-electron operators. Using a saddle point
approximation we get the Poisson equation for the SEC system. 
In section III we show how the periodical Hubbard-Anderson model appears
using 
the results of section II. In section IV we discuss the mixing parameter
entering the model, and in section V we conclude with a discussion. 

\section{The Poisson Equation in the System with SEC}
Here we reformulate the derivation given in Ref.~\cite{sand_fil}, using
the same ideology of separating the electrons in core and valence
states. 
The zero level Hamiltonian is the one considered in
Ref.~\cite{sand_fil}, 
but within a non-orthogonal basis set and, what is more important,
here we avoid the slave-boson technique in favor of the diagram
technique from
the atomic limit developed in Ref.~\cite{DT,II}.
Let us consider an ion which has $n$ \f-electrons 
in the ground state. Then, only the transitions 
$\Gamma_{n}\rightarrow \Gamma_{n\pm 1}$ 
will be allowed in the spectrum of excitations while all 
other transitions like $\Gamma_{n}\rightarrow \Gamma_{n\pm 2},\Gamma 
_{n\pm 3}$, involving a larger number of electrons, will be strongly 
suppressed by the large energy separation between these states. If the 
energy of 
the atomic-like transition $\Delta_{2}\equiv E_{\Gamma
}^{(n+1)}-E_{\Gamma
^{\prime}}^{(n)}$ between an $(n+1)$ and $n$-electron state, $\Gamma$
and $\Gamma^{\prime}$, of the $f$-ion is much higher than the Fermi
energy, $\varepsilon_{F}$, the number of $f$-electrons in the 
ion will be fixed. Indeed,
in this limit this upper ''single-electron'' level is empty while the
lower
one, even if it forms a band, will be fully filled. In the rare earth
elements
the populated part of the $f$-spectral density corresponding to the
transitions $\Delta _{1}\equiv E_{(n)}-E_{(n-1)}$ is much below 
$\varepsilon _{F}$
(except for Ce, and perhaps Sm).
It can be also much below the bottom of the conduction-electron bands.
When
an orbital has such a low energy, the mixing interaction, as well as
overlap 
between these core-like levels and conduction electrons are 
negligible~\cite{olof}. This physical picture exactly corresponds to the type
of \emph{ab initio} calculation where the $f$-electrons are kept in the core. 
One can use the many-electron functions for the description of the ground 
state of an ion from DFT-LDA-based calculations too. 
All electrons in this case experience the 
same potential. As has been shown in Ref.~\cite{II}, this picture is 
valid when the energy of the upper transition, 
$\Delta_{2}=E_{(n+1)}-E_n$, is much above 
Fermi energy. However, the photo-electron spectroscopy 
experiments show that even in rare earth elements, for which 
this picture seems to be most appropriate, the level $\Delta _{2}$ 
is sometimes 
only slightly above $\varepsilon_{F}$~\cite{baer}. Therefore, due to
mixing 
interaction and, possibly, hopping, a band with mixed $f$- and 
conduction electron states~\cite{urbansqrt} can be formed. 
As discussed in details 
in Ref.~\cite{II}, this leads to shifts of spectral weights from integer
values and a 
violation of the single-electron picture. These spectral weights,
therefore,
must enter the expression for the charge density in the Poisson
equation.
Besides, they control the strength of the mixing and hopping.
Let us derive the Poisson equation which contains the information about
these spectral weights and is valid not only for zero temperature. Here,
we 
will not consider the formation of phonons
and scattering processes which involve them, nor plasmons.

\subsection{The Hamiltonian in a Non-Orthogonal Basis Set}
An orthogonalization procedure of the wave-functions belonging to
different ions leads to a coupling of the states. This  makes it 
difficult to separate the strongest
single-site interactions. Therefore, the local strong interactions
between $f$-electrons can most easily be taken into account in the
\emph{non-orthogonal}
site representation. For this reason we will, to some extent, use the
technique developed previously~\cite{DT} (below referred to as I). The
delocalized electrons are treated within the weak-coupling perturbation
theory (WCPT), while the localized (or semi-localized) within the
strong-coupling theory (SCPT), see I.
In order to introduce, for the $f$-operators (and other core electrons),
the many-electron representation we rewrite the field
operator, $\hat{\psi}_{\sigma }(\rr)$, in the $jL$-representation 
\begin{eqnarray}
\hat{\psi}_{\sigma }(\rr) &=&\int \dd\qq\,e^{-i\qq\cdot\rr}
\phi _{jL}(\rr)a_{jL}, \label{psi:eq} \\
a_{jL} &=&[(1-\delta _{L,\mu })c_{jL}+\delta _{L,\mu }f_{j\mu }]. 
\label{two_class:eq}
\end{eqnarray}
Here, $j\equiv R_{j}$ is the site, $L\equiv (l,m_{l},s=1/2,\sigma )$,
$l$ is 
the orbital moment, $m_{l}$ is its projection to the $z$-axis, $s$ is
electron spin and $\sigma $ its projection to the same axis $\mu$ indicates 
localized electrons. In
Eq.(\ref{two_class:eq}) we 
have separated all electrons to two classes: core electrons, $f_{j\mu}$,
which either remain fully localized in solids, or only partly
delocalized,
and delocalized electrons, $c_{jL}$, which will be described in $\kk$-space
in
regular crystals. Since the essential part of the work to be done
concerns 
the localized electrons, it is reasonable to formulate the approach
in the site representation first. The basis functions $\phi _{jL}(\rr)$
are in general not
orthogonal to each other, 
\begin{equation}
\int \dd\rr\phi _{jL}^{\ast }(\rr)\phi^{\vphantom{\ast}}_{j^{\prime
}L^{\prime
}}(\rr)={\cal O}_{jL,j^{\prime }L^{\prime }}.
\end{equation}
therefore, from $\{\psi _{\sigma }(\rr),\psi _{\sigma ^{\prime
}}^{\dagger
}(\rr^{\prime })\}=\delta (\rr-\rr^{\prime })\delta _{\sigma \sigma
^{\prime
}}$,
we see that 
\begin{equation}
\{a^{\vphantom{\dagger}}_{jL},a_{j^{\prime }L^{\prime }}^{\dagger }\}
={\cal O}_{jL,j^{\prime}L^{\prime}}^{-1},
\end{equation}
where ${\cal O}_{jL,j^{\prime }L^{\prime }}^{-1}$ is the $(jL,j^{\prime
}L^{\prime
})$-matrix element of the inverse of the overlap matrix $\mathbf{O}$.

The full Hamiltonian is 
\begin{eqnarray}
&&\hspace*{-5mm}\HH =\int \dd\rr\psi _{\sigma }^{\dagger }(\rr)\left[
\frac{p^{2}}{2m}-\sum_{j}%
\frac{Z_{j}e^{2}}{|\rr-\RR_{j}|}-C_{\infty }\right] 
\psi^{\vphantom{\dagger}} _{\sigma
}(\rr)+\HH_{nn} 
\nonumber \\
&&+\frac{1}{2}\int \dd\rr\dd\rr^{\prime }\psi _{\sigma }^{\dagger }(\rr)
\psi^{\vphantom{\dagger}}_{\sigma
}(\rr)v(\rr-\rr^{\prime })\psi _{\sigma ^{\prime }}^{\dagger
}(\rr^{\prime
})\psi^{\vphantom{\dagger}}_{\sigma ^{\prime }}(\rr^{\prime }), 
\label{ham1:eq} \\
&&\HH_{nn} =\sum_{j}\frac{Z_{i}Z_{j}e^{2}}{|\RR_{i}-\RR_{j}|},
\end{eqnarray}
where $C_{\infty }$ is the infinite constant $\int
\dd\rr\dd\rr^{\prime
}\delta \rr-\rr^{\prime })v(\rr-\rr^{\prime })\sum_{\sigma ^{\prime
}}\delta
_{\sigma \sigma ^{\prime }}$ which arose when we transformed 
$\psi _{1}^{\dagger }\psi _{2}^{\dagger }
v(1,2)\psi^{\vphantom{\dagger}}_{2}\psi^{\vphantom{\dagger}}_{1}$ into
$\psi
_{1}^{\dagger }\psi^{\vphantom{\dagger}}_{1}v(1,2)\psi_{2}^{\dagger }
\psi^{\vphantom{\dagger}}_{2}$. We omit
this constant below, since it does not influence the physics. 
Let us rewrite the Hamiltonian in the representation using the functions
$\phi_{jL}(\rr)$
(which can also be defined in different ways and we shall discuss it
later).
Using the expansion in Eq.(\ref{psi:eq}) inserted into Eq.(\ref{ham1:eq}) 
gives 
\begin{eqnarray}
&&\hspace{-5mm}
\HH=\HH_{nn}+%
\sum_{j_{2}L_{2},j_{3}L_{3}}h_{j_{2}L_{2},j_{3}L_{3}}^{0}a_{j_{2}L_{2}}^{%
\dagger }a^{\vphantom{\dagger}}_{j_{3}L_{3}} \nonumber \\
&&+\left[ \frac{1}{2}\sum_{\{jL%
\}}v^{\vphantom{\dagger}}_{j_{2}L_{2},j_{3}L_{3},j_{4}L_{4},j_{5}L_{5}}
a_{j_{2}L_{2}}^{\dagger
}a^{\vphantom{\dagger}}_{j_{5}L_{5}}a_{j_{3}L_{3}}^{\dagger }
a^{\vphantom{\dagger}}_{j_{4}L_{4}}\right] .
\end{eqnarray}
Here 
\begin{equation}
h_{j_{2}L_{2},j_{3}L_{3}}^{0}\equiv (j_{2}L_{2}|\left[
\frac{p^{2}}{2m}%
-\sum_{j}\frac{Z_{j}e^{2}}{|\rr-\RR_{j}|}\right] |j_{3}L_{3}),
\end{equation}
Now we assume that the
nuclei are in fixed positions and separate the part of the Hamiltonian
which
contains on-site interactions between electrons that are treated as 
core electrons
\begin{eqnarray}
\HH^{ion} &=&\sum_{j}\HH_{j}^{ion}=\sum_{j}\left\{ \sum_{\mu _{2},\mu
_{3}}h_{j\mu _{2},,j\mu _{3}}^{0}f_{j\mu _{2}}^{\dagger }
f^{\vphantom{\dagger}}_{j\mu_{3}}\right.
\nonumber \\
&&+\left.  \frac{1}{2}\sum_{\{\mu \}}
v^{\vphantom{\dagger}}_{j\mu _{2},j\mu _{3},j\mu
_{4},j\mu _{5}}a_{j\mu _{2}}^{\dagger }
a^{\vphantom{\dagger}}_{j\mu _{5}}a_{j\mu_{3}}^{\dagger
}a^{\vphantom{\dagger}}_{j\mu _{4}} \right\} .
\end{eqnarray}
The single-site part of the problem and the rest will be treated in
different approximations. We want to use Hubbard operators that 
are usually introduced
in such a way that they diagonalize the single-ion Hamiltonian 
\begin{eqnarray}
H_{j}^{ion}|j,\Gamma _{n}\rangle &=&E_{j\Gamma _{n}}|j,\Gamma
_{n}\rangle , \\
X_{j}^{\Gamma _{n}\Gamma _{m}} &\equiv &|j,\Gamma _{n}\rangle
\langle j,\Gamma _{m}|.
\end{eqnarray}
Let us now discuss how to construct them. 

\subsection{The Hubbard Operators in Terms of Fermions}
We are not able to diagonalize the full Hamiltonian exactly, unless for
certain model
calculations, and we have to use some approximation. In order to
ensure that the ground-state wave function fulfills Hunds 
rules, we will follow the technique by Irkhin~\cite{irkhin}, who
translated
the Racah technique, used in atomic
spectroscopy~\cite{sobelman} for the wave functions, 
into the operator language. Although this issue has
been discussed in Refs.~\cite{irkhiny,DT}, we find the definition used
not fully
satisfactory and in need of a slight modification. For this reason we
discuss the definition of the many-electron operators and, correspondingly,
the connection between the Hubbard operators and the many-electron
creation and annihilation 
operators and the modification needed. The creation operator for a group
of $n$ equivalent electrons (say, in an $f$-shell) in the many-electron
state
$|\Gamma _{n}\rangle =A_{\Gamma _{n}}^{\dagger }|0\rangle $ in Irkhin's 
definition has the form 
\begin{equation}
\label{irkhin:eq}
A_{\Gamma _{n}}^{\dagger }=\frac{1}{\sqrt{n}}\sum_{\mu ,\Gamma
_{n-1}}G_{\Gamma _{n-1}}^{\Gamma _{n}}C_{\Gamma _{n-1},\mu }^{\Gamma
_{n}}f_{\mu }^{\dagger }A_{\Gamma _{n-1}}^{\dagger }.
\end{equation}
Here $G_{\Gamma _{n-1}}^{\Gamma _{n}}$ are the fractional parentage
coefficients, which do not depend on the momentum projections (if 
$n\leq 2$,  
$G=1$ and the squared coefficient, $(G_{\Gamma _{n-1}}^{\Gamma
_{n}})^{2}$ ,
measures the fracture of the state $|\Gamma _{n-1}\rangle $ in 
$|\Gamma_{n}\rangle $); $C_{\Gamma _{n-1},\mu }^{\Gamma _{n}}$ are the 
Clebsch-Gordan coefficients, 
\begin{equation}
C_{\Gamma _{n-1},\mu }^{\Gamma _{n}}\equiv
C_{L_{n}M_{L}^{(n-1)},lm_{l}}^{L_{n}M_{L}^{(n)}}C_{S_{n-1}M_{S}^{(n-1)},1/2,%
\sigma }^{S_{n}M_{S}^{(n)}},
\end{equation}
where $L_{n},M_{L}^{(n)},S_{n},M_{S}^{(n)}$ are the orbital moment, its
projection, spin moment and its projection for the $n$-electron
configuration
$|\Gamma _{n}\rangle $. In order to be able to calculate the commutation
relations between the conduction electrons and the Hubbard operators 
(H-operators) as well as between the H-operators themselves, we have to
express them in terms of Fermion operators. We have to provide the
correct
commutation relations for H-operators, belonging to the same site 
\begin{equation}
\lbrack X^{\gamma \Gamma },X^{\Lambda \chi }]_{\pm }=\delta ^{\Gamma
\Lambda
}X^{\gamma \chi }\pm \delta ^{\chi \gamma }X^{\Lambda \Gamma }.
\end{equation}
If we try to define an H-operator in the form of a product of the
operators $A$,i.e.\ $X^{\gamma\Gamma}=A_{\gamma}^{\dagger}A_{\Gamma}$, 
we should get
zero if we multiply by $X^{\gamma\Gamma_{n}}X^{\Lambda_{m}\chi}$ for
$n\neq m$. 
This should be provided by the fact, that for
the Fermion operators $%
f^{2}=(f^{\dagger})^{2}=0$. However, the product $A_{\Gamma
_{n}}A_{\Lambda
_{m}}^{\dagger }\neq 0$, if $n<m$, because the operators $A_{\Gamma
_{n}}$ do
not contain information about non-filled orbitals of the shell. The
operators $A_{\Gamma _{n}}$ do not provide the orthogonality of the
states
with different number of electrons in shell. The recipe suggested in
Ref.~\cite{irkhiny} is to define new operators as follows 
\begin{equation}
\tilde{A}_{\Gamma _{n}}^{\dagger }=A_{\Gamma _{n}}^{\dagger }\prod_{\mu
}(1-\hat{n}_{\mu }),\;\tilde{A}_{\Gamma _{n}}=\prod_{\mu
}(1-\hat{n}_{\mu
})A_{\Gamma _{n}}.
\end{equation}
The product should run over all non-filled orbitals. Let us consider,
for example, the two-electron state composed of \f-states, 
$|\Gamma_{2}\rangle
=|L=5,M_{L}^{(2)}=4,S=1,M_{S}^{(2)}=0\rangle$. Then, the operator 
\begin{equation}
A_{\Gamma _{n}}^{\dagger }=\sum_{m_{1}m_{2}\sigma _{1}\sigma
_{2}}C_{3m_{1},3m_{2}}^{5,4}C_{1/2,\sigma _{1};1/2,\sigma
_{2}}^{1,0}f_{3m_{1};1/2,\sigma _{1}}^{\dagger }f_{3m_{2};1/2,\sigma
_{2}}^{\dagger }
\end{equation}
combines three terms (for briefness below we omit indices $l=3$ and spin
$%
s=1/2$ : 
\begin{equation}
A_{\Gamma _{2}}^{\dagger }=\alpha \lbrack f_{3\uparrow }^{\dagger
}f_{1\downarrow }^{\dagger }+\gamma f_{2\uparrow }^{\dagger
}f_{2\downarrow
}^{\dagger }+f_{1\uparrow }^{\dagger }f_{3\downarrow }^{\dagger }]
\end{equation}
with $\alpha =\sqrt{5/253},\gamma =27/\sqrt{15}$. 
Now, we have to multiply this by the product of the factors
$(1-\hat{n}%
_{\mu })$, where $\mu $ runs over all empty orbitals. From this a
problem is
apparent: the different terms of the combinations of Clebsch-Gordan
coefficients involve
\emph{%
different} orbitals and, therefore, it is impossible to choose a single
factor which includes all empty orbitals for each term in the sum.
Therefore, \emph{each term} of the sum must be supplemented with its own
factor. In this particular example the new many-electron operator should
be
defined as follows 
\begin{eqnarray}
\tilde{A}_{\Gamma _{2}}^{\dagger } &=&\alpha \left[ f_{3\uparrow
}^{\dagger
}f_{1\downarrow }^{\dagger }\prod_{m_{1}\neq 3}(1-\hat{n}_{m_{1}\uparrow
})\prod_{m_{2}\neq 1}(1-\hat{n}_{m_{2}\downarrow })+\gamma f_{2\uparrow
}^{\dagger }f_{2\downarrow }^{\dagger }\prod_{m_{1}\neq 2}(1-\hat{n}%
_{m_{1}\uparrow })\prod_{m_{2}\neq 2}(1-\hat{n}_{m_{2}\downarrow
})\right. 
\nonumber \\
&&+\left. f_{1\uparrow }^{\dagger }f_{3\downarrow }^{\dagger
}\prod_{m_{1}\neq 1}(1-\hat{n}_{m_{1}\uparrow })\prod_{m_{2}\neq
3}(1-\hat{n}%
_{m_{2}\downarrow })\right] \equiv \alpha \lbrack \tilde{A}_{\Gamma
_{2}(3\uparrow ,1\downarrow )}^{\dagger }+\gamma \tilde{A}_{\Gamma
_{2}(2\uparrow ,2\downarrow )}^{\dagger }+\tilde{A}_{\Gamma
_{2}(1\uparrow
,3\downarrow )}^{\dagger }].  \nonumber \\
&& \label{tot_prod:eq}
\end{eqnarray}
Obviously, in a general case, when we construct the operator $\tilde{A}%
_{\Gamma _{2}}^{\dagger }$, \emph{each term} in the sum should be
multiplied by the product of the factors $(1-\hat{n}_{m\sigma })$ 
corresponding to this term,  
where the set of $\{m\sigma \}$ includes only those orbitals which are
not included in the product of the $f$-operators in the corresponding 
term of the
Clebsch-Gordan sum. Let us denote this product $\bar{\Pi}$, where the
bar over $\Pi $ means that it contains only complementary orbitals. In I
we used the \emph{orbital representation,} where each H-operator
contains
only one term, containing $n$ creation $f$-operators for the
$n$-electron
state which is multiplied by the product $\bar{\Pi}$. We we will call it
\emph{elementary operator}. In the case of Eq.(\ref{tot_prod:eq}) 
these operators are
$\tilde{A}_{\Gamma _{2}(3\uparrow ,1\downarrow )}^{\dagger
},\;\tilde{A}%
_{\Gamma _{2}(2\uparrow ,2\downarrow )}^{\dagger },\;\tilde{A}_{\Gamma
_{2}(1\uparrow ,3\downarrow )}^{\dagger }$. We can, therefore, make the
statement that any operator of a state in a central field (i.e.\ of the
Clebsch-Gordan combination type) or in a crystal field, can be
represented
as a sum of elementary operators with the coefficients, which are
dictated by
the symmetry of the surrounding of the ion. 
Since the construction of the state in central
field within the Racah technique is recursive, i.e.\ the $n$-electron
state is composed of $(n-1)$-electron states and one-electron states,
the $(n-1)$-one is made of the combination of $(n-2)$-and one-electron
states, 
and so on, this statement needs proof. Let us start with a many-electron
operator, describing one localized electron in orbital $1$. The
electron state has the form $f_{1}^{\dagger }\prod_{\nu \neq
1}(1-\hat{n}%
_{\nu })$. In order to construct the state, which contain two electrons
localized in the states 1 and 2, we have to multiply this operator by $%
f_{2}^{\dagger }$ from the left-hand side. Since $f_{2}^{\dagger
}(1-\hat{n}%
_{2})=f_{2}^{\dagger }$, all extra factors $(1-\hat{n}_{\nu })$ are
automatically projected out and we are back at Eq.(\ref{irkhin:eq}). 
Therefore, each
step to a higher number electron operator will be started again with
Eq.(\ref{irkhin:eq}). This means that the factors $\bar{\Pi}$ should 
be added in each
term of the sum in the last step only. Thus, the Hubbard operator can be
written in the form 
\begin{equation}
X^{\gamma \Gamma }=\tilde{A}_{\gamma }^{\dagger }\tilde{A}_{\Gamma },
\end{equation}
where each elementary operator entering the sum for the operator
$\tilde{A%
}_{\gamma }^{\dagger }$ contains the projecting product
$\bar{\Pi}_{\gamma
}^{(i)}$. The same is valid for $\tilde{A}_{\Gamma }$ and
$\bar{\Pi}_{\Gamma}^{(j)}$.
Then, we can represent each $f_{j\mu }$-operator in any place where we
meet it, particularly, in the Hamiltonian, in terms of H-operators $%
X_{j}^{a}\equiv X_{j}^{a(\gamma ,\Gamma )}\equiv X_{j}^{[\gamma ,\Gamma
]}$: 
\begin{equation}
f_{j\mu }\equiv (f_{\mu })^{a}X_{j}^{a}.
\end{equation}
Here, repeated indices are summed over. 
The H-operators can also be written in the Hubbard
form $X_{j}^{[\gamma ,\Gamma ]}\equiv |\RR_{j},n,\gamma \rangle \langle
\RR_{j},n+1,\Gamma |$ or in terms of products of Fermion operators as
discussed above (here $|\RR_{j},n,\gamma \rangle $ and
$|\RR_{j},n+1,\Gamma
\rangle $ are many-electron $n$- and $(n+1)$-particle states). Using the
definition of the many-electron operators and the Hubbard operators in
terms of Fermion operators one can calculate all commutation
relations between the conduction electrons operators and the Hubbard
operators~\cite{DT},
\begin{equation}
\{c^{\vphantom{\bar{a}}}_{jL},X_{j^{\prime }}^{\bar{a}}\}=
O_{jL,j^{\prime }\nu }^{-1}f_{\nu
}^{b}\varepsilon _{\xi }^{b\bar{a}}Z_{j^{\prime }}^{\xi }.
\end{equation}
Here, $a,b$ denote the Fermi-like transitions from the $n$- to
($n+1$)-electron
state ($a=a(\Gamma _{n},\Gamma _{n+1})$), $\bar{a}$ denotes the inverse
transition ($\bar{a}=\bar{a}(\Gamma _{n+1},\Gamma _{n})$), 
$\xi =\xi (\Gamma
_{n},\Gamma_{n}^{\prime})$ 
and $\varepsilon_{\xi}^{b\bar{a}}$ are the structure constants of the
algebra for the Hubbard operators, 
\begin{equation}
\{X_{j}^{b},X_{j^{\prime }}^{\bar{a}}\}=\varepsilon _{\xi }^{b\bar{a}%
}Z_{j^{\prime }}^{\xi },\;\{X_{j}^{b},Z_{j^{\prime }}^{\xi
}\}=\varepsilon
_{a}^{b\xi }X_{j^{\prime }}^{a},\;\{X_{j}^{\bar{b}},Z_{j^{\prime }}^{\xi
}\}=\varepsilon _{\bar{a}}^{\bar{b}\xi }X_{j^{\prime }}^{\bar{a}}.
\end{equation}
The summation over repeating indices is implied. 
A Bose-like transition is denoted as $Z^{\xi}$ and a
diagonal Bose-like
operator is denoted as $h_{j}^{\Gamma }$, $h_{j}^{\Gamma }\equiv
Z_{j}^{\xi
(\Gamma ,\Gamma )}$.

\subsection{The Field for the Electric Potential and the Poisson
Equation}
Let us introduce the Fourier-component of the charge density for the 
nuclei in the point 
$\RR_{\alpha }$ with the charge $Z_{\alpha}$: 
\begin{equation}
\hat{\rho}_{n}(\qq)=\sum_{j}Z_{j}e^{i\qq\cdot\RR_{j}}.
\end{equation}
Then, the interaction between the nuclei is 
\begin{equation}
\HH_{nn}^{coul}=\frac{1}{2}\sum_{jj^{\prime}}\frac{Z_{j}Z_{j^{\prime
}}e^{2}}{%
|\RR_{j}-\RR_{j^{\prime }}|}=\int \dd\qq\hat{\rho}_{i}(\qq)\frac{4\pi
e^{2}}{q^{2}}%
\hat{\rho}_{i}(-\qq).
\end{equation}
The interaction between electrons and nuclei,
\begin{eqnarray}
\HH_{en}^{coul}&=&-\sum_{j}\int
\dd\rr\hat{\rho}_{e}(\rr)\frac{Z_{j}e^{2}}{|\RR_{j}-\rr|}\nonumber \\
&&=-\int \dd\qq\hat{\rho}_{e}(\qq)\frac{4\pi
e^{2}}{q^{2}}\hat{\rho}_{n}(-\qq),
\end{eqnarray}
contains the electrons which are localized to the same nucleus, or to
other ones, 
and those which are in a mixed state. The
interaction of the localized electrons with their own nucleus is 
\begin{eqnarray}
&&
-\sum_{j}\int \dd\rr\frac{Z_{j}e^{2}}{|\rr-\RR_{j}|}\phi _{\nu }^{\ast
}(\rr-\RR_{j})\phi _{\mu }(\rr-\RR_{j})(f_{\nu }^{\dagger }
f^{\vphantom{\dagger}}_{\mu })^{\xi
}Z_{j}^{\xi }\nonumber \\ 
&&=-\int \dd\qq\frac{4\pi
e^{2}}{q^{2}}\sum_{j}Z_{j}S_{j}^{\xi
}(\qq)Z_{j}^{\xi },
\end{eqnarray}
where $S_{j}^{\xi }$ is the form-factor of the ion, which takes into
account
the contribution of the orbitals $\nu $ and $\mu $ into the transition
$\xi$ 
\begin{equation}
{\cal O}_{j}^{\xi }(q)=\int dre^{iqr}\phi _{j\nu }^{\ast }(r)\phi _{j\mu
}(r)(f_{\nu }^{\dagger }f^{\vphantom{\dagger}}_{\mu })^{\xi }
\end{equation}
and the index $j$ in $\phi_{j\mu}(\rr)$ denotes affiliation of this
function
to the ion on $\RR_{j}$. This term is included to the Hamiltonian for 
the ion in this point. The term 
\begin{eqnarray}
&&-\sum_{jn}\int \dd\rr\frac{Z_{j}e^{2}}{|\rr-\RR_{j}|}(1-\delta
_{jn})\phi
_{\nu
}^{\ast }(\rr-\RR_{n})\nonumber \\
&&\times\phi _{\mu }(\rr-\RR_{n})(f_{\nu }^{\dagger }
f^{\vphantom{\dagger}}_{\mu })^{\xi
}Z_{n}^{\xi }  \nonumber \\
&&=-\int \dd\qq\frac{4\pi e^{2}}{q^{2}}\sum_{j\neq n}
[Z_{j}e^{-i\qq\cdot\RR_{j}} {\cal O}_{n}^{\xi }(\qq)Z_{n}^{\xi
}e^{i\qq\cdot\RR_{n}}]
\end{eqnarray}
describes the interaction of localized electrons belonging to one ion
with a 
nucleus of another ion. 
Then, the operator of an effective density of ions can
be written as 
\begin{equation}
\hat{\rho}_{i}(\qq)=\sum_{j}[Z_{j}-\sum_{\xi }{\cal O}_{j}^{\xi
}(\qq)Z_{j}^{\xi
}]e^{i\qq\cdot\RR_{j}}
\end{equation}
and the interaction between all nuclei which are screened by their
electrons is 
\begin{equation}
\frac{1}{2}\int \dd\qq\hat{\rho}_{i}(\qq)\frac{4\pi
e^{2}}{q^{2}}\hat{\rho}%
_{i}(-\qq)-\HH_{D}.
\end{equation}
Here, $\HH_{D}$ takes into account the terms which are double counted in
the
first term, since there is no interaction of the ion with itself  
\begin{eqnarray}
&&\hspace*{-5mm}
\HH_{D}=\frac{1}{2}\sum_{j}\int \dd\qq[Z_{j}-\sum_{\xi }{\cal O}_{j}^{\xi
}(\qq)Z_{j}^{\xi
}]\frac{4\pi e^{2}}{q^{2}}[Z_{j}\nonumber \\
&&-\sum_{\xi }{\cal O}_{j}^{\xi }(\qq)Z_{j}^{\xi
}].
\end{eqnarray}
The terms of this interaction at small $\qq$, as well as $\HH_{D}$
itself,
diverge. This is the standard problem of screening.

All other electrons belong either to the class of the transitions
between 
different ions, or to a mixed state between conduction-electron and
localized
electrons, or to the conduction electrons. The operator of the charge
density, 
$\hat{\rho}_{\rr}(\qq)$, of these remaining electrons can be written as
follows 
\begin{eqnarray} 
&&\hspace*{-2mm}\hat{\rho}_{\rr}(\qq) =\sum_{_{jLj^{\prime }L^{\prime }}}
{\cal O}^{\vphantom{\dagger}}_{jLj^{\prime
}L^{\prime }}(\qq)c_{jL}^{\dagger }
c^{\vphantom{\dagger}}_{j^{\prime }L^{\prime }}  \nonumber \\
&&+\sum_{_{jLj^{\prime }L^{\prime }}}
{\cal O}^{\vphantom{\dagger}}_{jLj^{\prime }\mu ^{\prime
}}(\qq)(f^{\vphantom{\dagger}}_{\mu ^{\prime }})^{a}c_{jL}^{\dagger }
X_{j^{\prime}}^{a}+{\cal O}^{\vphantom{\dagger}}_{j\mu
j^{\prime }L^{\prime}}(\qq) (f_{\mu}^{\dagger })^{\bar{a}%
}X_{j}^{\bar{a}}c_{j'L'}  \nonumber \\ 
&&+(1-\delta^{\vphantom{\dagger}}_{jj^{\prime }})
{\cal O}^{\vphantom{\dagger}}_{j\mu j^{\prime }\mu^{\prime}}(\qq)(f_{\mu
}^{\dagger })^{\bar{a}}(f^{\vphantom{\dagger}}_{\mu ^{\prime
}})^{b}X_{j}^{\bar{a}}X_{j^{\prime
}}^{b}
\label{fourr:eq}
\end{eqnarray}
The matrices ${\cal O}(\qq)$ are defined as 
\begin{equation}
{\cal O}_{jLj^{\prime }L^{\prime }}(\qq)=\int \dd\rr e^{i\qq\cdot\rr}\phi
_{jL}^{\ast}(\rr)\phi^{\vphantom{\dagger}}
_{j^{\prime }L^{\prime }}(\rr).
\end{equation}
Thus, we obtain the Hamiltonian in the many-electron representation 
\begin{equation}
\HH=\HH_{0}+(\HH^{coul}+T),
\end{equation}
where $T$ is the kinetic energy, and the zero Hamiltonian is 
\begin{equation}
\HH_{0}=\HH_{0}^{X}+\HH_{0}^{c}=\sum_{j\Gamma }
E^{\vphantom{\dagger}}_{j\Gamma }h_{j}^{\Gamma}+\sum
h_{jL,j^{\prime }L^{\prime }}^{0c}c_{jL}^{\dagger }
c^{\vphantom{\dagger}}_{j^{\prime}L^{\prime}}.
\end{equation}
Here $\HH_{0}^{X}$ describes the electrons treated as core electrons
when the interaction between the ions and all other electrons is
switched off  
\begin{eqnarray}
&&\hspace*{-5mm}
E_{j\Gamma }=(j,\Gamma |\sum_{i\in
\RR_{j}}\frac{p_{i}^{2}}{2m}-\sum_{i\in
\RR_{j}}\frac{Z_{j}e^{2}}{|\RR_{j}-\rr_{i}|}\nonumber \\
&&+\frac{1}{2}\sum_{i,i^{\prime
}\in
\RR_{j}}\frac{Z_{j}e^{2}}{|\rr_{i}-\rr_{i^{\prime }}|}|j,\Gamma ),
\end{eqnarray}
and the function $|j,\Gamma )$ is a many-electron wave-function.
The Hamiltonian of the conduction electrons, $\HH_{0}^{c}$,
describes the electrons in the field of the unscreened nuclei. This will
be 
supplemented with the self-consistent field from the electrons. The
kinetic
energy of the electrons which belong to one site is taken into account,
while the part $T$ of it, which is proportional to
$\hat{\rho}_{\rr}(\qq)$, is
not. We will treat this part of the kinetic energy as a perturbation.
Thus,
the full Coulomb interaction can be written as follows 
\begin{eqnarray}
&&\hspace*{-5mm}\frac{1}{2}\int
\dd\qq[\hat{\rho}_{i}(\qq)-\hat{\rho}_{\rr}(\qq)]\frac{4\pi
e^{2}}{%
q^{2}}[\hat{\rho}_{i}(\qq)-\hat{\rho}_{\rr}(\qq)]\nonumber \\
&&\equiv \frac{1}{2}\int
\dd\qq\hat{\rho%
}_{\qq}\frac{4\pi e^{2}}{q^{2}}\hat{\rho}_{\qq}.
\end{eqnarray}

The partition function, written in the standard form, is 
\begin{eqnarray}
Z &=&Tr\exp [-\beta (\HH-\mu N)]\nonumber \\
&=&Z_{0}\frac{1}{Z_{0}}Tr\{\exp[-\beta
\HH_{0}]{\cal T}_{\tau }\exp[-\int_{0}^{\beta }\dd\tau \HH_{int}(\tau )]\}
\nonumber \\
&\equiv &Z_{0}\cdot \langle {\cal T}_{\tau }\exp [-\int_{0}^{\beta }d\tau
\HH_{int}(\tau )]\rangle ^{(0)} 
\end{eqnarray}
Let us write the part $\HH_{\rho \rho }^{coul}$ of $\HH_{int}$ in the
Fourier
transformed form, Eq.(\ref{fourr:eq}). 
Then, we introduce the Gaussian functional integral 
\begin{equation}
1=\int D\varphi _{\qq}(\tau )\exp [-\int_{0}^{\beta }\frac{\dd\tau
}{8\pi
}\int
\dd\qq\;q^{2}\varphi_{\qq}(\tau )\varphi_{-\qq}(\tau )]
\end{equation}
under the trace of the partition function. We make the shift 
\begin{equation}
\varphi _{\qq}(\tau )\rightarrow \varphi _{\qq}(\tau )+i\frac{4\pi
e}{q^{2}}\hat{%
\rho}_{\qq},
\end{equation}
in this Gaussian integral. This allows us to rewrite the Coulomb
interaction in
terms of interaction of electrons with the random field $\varphi
_{\qq}(\tau )$. This shift generates the term 
$(-\HH_{\rho \rho }^{coul})$ which
cancels the $\HH_{\rho \rho }^{coul}$ in the $\HH_{int}$ but adds the
term
\begin{equation}
-\frac{ie}{2}\int_{0}^{\beta }\dd\tau \int \dd\qq\;[\varphi _{\qq}(\tau
)\hat{\rho}_{-\qq}(\tau )+\hat{\rho}_{\qq}(\tau )\varphi _{-\qq}(\tau
)].
\end{equation}
Note that, although the $f$-orbitals of the same site are
orthogonalized,
the 
$\qq\neq 0$-components of the overlap matrix have non-zero values and,
therefore, non-diagonal transitions $\xi \neq \lbrack \Gamma ,\Gamma ]$
enter the expression for $\qq$-component of the charge density. Thus, we
have
to work with the following expression for the partition function  
\begin{eqnarray}
&&\hspace*{-5mm}
\frac{Z}{Z_{0}} =\int D\varphi _{\qq}(\tau )\exp
[-\int_{0}^{\beta }\dd\tau \int \dd\qq\;\frac{q^{2}}{8\pi }\varphi
_{\qq}(\tau)\varphi_{-\qq}(\tau)]  \nonumber \\
&&\times \exp [-\beta \mathcal{F}_{i}[\varphi _{\qq}(\tau ),\varphi
_{-\qq}(\tau )]], 
\end{eqnarray}
where
\begin{eqnarray}
&&\hspace*{-5mm}
\mathcal{F}_{i}[\varphi _{\qq}(\tau ),\varphi _{-\qq}(\tau )] 
\nonumber \\
&&\equiv -\frac{1}{
\beta}\ln \langle {\cal T}_{\tau }\exp [-\int_{0}^{\beta }\dd\tau \int
\dd\qq(T_{\qq}-\HH_{\rho }\nonumber \\
&&\hspace*{1cm}-\HH_{D})]\rangle ^{(0)}, \\
T_{\qq} &\equiv &\left[ \frac{\hat{p}^{2}}{2m}\right] _{\qq}\rho
_{\rr}(\qq) \\
\HH_{\rho } &=&\frac{1}{2}\int \dd\qq\;[(ie\varphi _{\qq}(\tau )-\mu
)\hat{\rho}%
_{-q}(\tau )+\hat{\rho}_{\qq}(\tau )\nonumber \\
&&\hspace*{1cm}\times(ie\varphi _{-\qq}(\tau )-\mu )],
\\
&&\HH_{i}^{0} =\HH_{0}^{c}+\HH_{0}^{X}. 
\end{eqnarray}
Here, we have used the fact that the fields $\varphi_{\qq}(\tau )$
commute with any operator. The function 
$\mathcal{F}$ is written in the form usually used for the cumulant
expansion.

Since both the mixing interaction and overlap matrices are non-zero, 
a part of the charge is in the mixed $cf$-states. Besides, the 
$f$-subsystem is described in terms of \emph{non-linear} $X$-operators.
For
these reasons we cannot describe the full contribution from the term 
$\mu\hat{N}$ in the zero Hamiltonian.
The field $\varphi_{\qq}(\tau)$ can be interpreted as the field for the
electrical potential which is generated by Coulomb interaction. If we
put 
$e=0$ the system does not have charged particles and, therefore, there
are no
contribution to the partition function from this field. At $e\neq 0$ in
a
non-homogeneous
system an average charge density is not equal to zero, and, therefore,
the
expectation value $\langle\varphi_{\qq}(\tau )\rangle_{\varphi}$ is
non-zero 
too. Taking the functional derivative of the free energy, we find
that the saddle-point approximation generates the Poisson equation for 
this field  
\begin{eqnarray}
\frac{\delta F}{\delta \varphi _{-\qq}(\tau )}&\equiv&\frac{\delta 
(-{\cal T}\ln
Z)}{%
\delta \varphi _{-\qq}(\tau )}\nonumber \\
&=&-\frac{q^{2}}{8\pi }\langle \varphi
_{\qq}(\tau
)\rangle _{\varphi }+\frac{ie}{2}\langle \hat{\rho}_{\qq}(\tau )\rangle
_{\varphi }=0.
\end{eqnarray}
The static part of the field of the electrical potential $\Phi(\rr)$ is
connected with the field $\langle \varphi\rangle $ as follows  
\begin{equation}
\Phi (\rr)\equiv \lim_{\tau \rightarrow -0}i\langle \varphi (\rr,\tau
)\rangle
_{\varphi }=i\langle \varphi (\rr,0)\rangle _{\varphi }.
\end{equation}
Thus, we have the Poisson equation for the electric field $\Phi(\rr)$, 
\begin{equation}
\nabla^{2}\Phi(\rr)=-4\pi \langle \hat{\rho}_{\qq}(0)\rangle _{\varphi},
\end{equation}
where the charge density is the difference between the densities of the
delocalized electrons and ''soft'' ion 
(where the polarization of the ion and excitations are allowed).

\section{The Self-Consistent Hubbard-Anderson Model}
Since the charge density, $\langle \hat{\rho}_{\qq}(0)\rangle
_{\varphi}$, contains expectation values $\langle c_{jL}^{\dagger
}c^{\vphantom{\dagger}}_{j^{\prime }L^{\prime }}\rangle $, 
$\langle c_{jL}^{\dagger
}X_{j^{\prime }}^{a}\rangle $, $\langle X_{j^{\prime }}^{\bar{a}
}c^{\vphantom{\dagger}}_{jL}\rangle $, 
$\langle X_{j}^{\bar{a}}X_{j^{\prime }}^{b}\rangle $,
we need expressions for them. Since the field $\varphi (\rr,\tau )$
contains the average static field and deviations of it 
\begin{equation}
\varphi (\rr,\tau )=\frac{1}{i}\Phi (\rr)+\delta \varphi (\rr,\tau ),
\end{equation}
it is reasonable to start with the approximation $i\varphi _{\qq}(\tau
)_{\varphi }\rightarrow i\langle \varphi _{\qq}(0)\rangle _{\varphi
}\rightarrow \Phi (\qq)$, and to neglect the fluctuations of this field.
Then, we
obtain the following saddle-point Hamiltonian 
\begin{eqnarray}
\tilde{\HH}_{0} &=&\int \dd\qq\;q^{2}\Phi (\qq)\Phi (-\qq)\nonumber \\
&&+\sum
(h_{jL,j^{\prime
}L^{\prime }}-\mu \,{\cal O}_{jL,j^{\prime }L^{\prime }})c_{jL}^{\dagger
}c^{\vphantom{\dagger}}_{j^{\prime }L^{\prime }}  \nonumber \\
&&+\sum (E_{j\Gamma }^{(n)}-n\mu )h_{j}^{\Gamma }, \\
\tilde{\HH}_{int} &=&T+\frac{e}{2}\int \dd\qq\;[\Phi
(\qq)\hat{\rho}_{-\qq}+\hat{%
\rho}_{\qq}\Phi (-\qq)]-\HH_{D}.
\end{eqnarray}
Here 
\begin{equation}
h_{jL,j^{\prime}L^{\prime}}=(jL|\frac{\hat{p}^{2}}{2m}+\Phi
(\rr)|j^{\prime}L^{\prime})
\end{equation}
is the frequency matrix of the conduction electrons in the
self-consistent
field $\Phi(\rr)$. The Hamiltonian $\mathbf{\tilde{h}}$ for them 
can be obtained if we make a transformation to the orthogonal variables
$\mathbf{\alpha}$, using the Cholesky decomposition for the 
overlap matrix $\mathbf{O}$: 
\begin{eqnarray}
\mathbf{c}^{\dagger }\mathbf{(h-\mu \,O)c} &=&\mathbf{c}^{\dagger
}[\mathbf{%
\bar{Z}(\bar{Z}}^{-1}\mathbf{hZ}^{-1})\mathbf{Z-}\mu
\mathbf{\,\bar{Z}Z]c=} 
\nonumber \\
(\mathbf{c}^{\dagger }\mathbf{\bar{Z})[\tilde{h}-}\mu \mathbf{I](Zc)}
&=&%
\mathbf{\alpha }^{\dagger }\mathbf{[\tilde{h}-}\mu \mathbf{I]\alpha .}
\end{eqnarray}
Let us now insert into $\tilde{\HH}_{int}$ the expression for
$\hat{\rho}_{\qq}$ 
in the $jL$-representation. Taking into account that 
\begin{eqnarray}
&&\hspace*{-5mm}
\frac{e}{2}\int \dd\qq\;\Phi (\qq){\cal O}_{jL,j^{\prime }L^{\prime
}}(\qq)=\frac{e}{2}%
\int \dd\rr\;\phi _{jL}^{\ast }(\rr)\Phi(\rr)
\phi^{\vphantom{\ast}}_{j^{\prime }L^{\prime
}}(\rr)\nonumber \\ 
&&\hspace*{2cm}\equiv V_{jL,j^{\prime }L^{\prime }}
\end{eqnarray}
are matrix elements of the self-consistent Coulomb field, we find that
$\tilde{\HH}$ actually gives the periodic Hubbard-Anderson Hamiltonian
\begin{eqnarray}
\tilde{\HH} &=&\int \dd\qq\;q^{2}\Phi (\qq)\Phi (-\qq)\nonumber \\
&&+\sum
(h^{LDA}-\mu \,{\cal O})^{\vphantom{\dagger}}_{jL,j^{\prime
}L^{\prime }}
c_{jL}^{\dagger
}c^{\vphantom{\dagger}}_{j^{\prime }L^{\prime }}  \nonumber \\
&&+\sum (\tilde{E}_{j\Gamma }^{(n)}-n\mu )h_{j}^{\Gamma }\nonumber \\
&&+\sum [V_{j\mu
,j\mu ^{\prime }}(f_{\mu }^{\dagger }
f^{\vphantom{\dagger}}_{\mu ^{\prime }})^{\xi \neq
\lbrack
\Gamma ,\Gamma ]}Z^{\xi }+h.c.] \nonumber \\
&&+\sum (\frac{p^{2}}{2m}+V-\mu {\cal O})^{\vphantom{\dagger}}_
{jL,j^{\prime }\mu ^{\prime }}(f^{\vphantom{\dagger}}_{\mu
^{\prime }})^{a}c_{jL}^{\dagger}X_{j^{\prime
}}^{a}\nonumber \\
&&+(\frac{p^{2}}{2m}+V-\mu
{\cal O})^{\vphantom{\dagger}}_{j\mu ,j^{\prime}\mu^{\prime}}(f_{\mu
}^{\dagger
})^{\bar{a}}X_{j}^{%
\bar{a}}c_{j^{\prime }L}] \nonumber \\
&&+\sum (1-\delta _{jj^{\prime }})(\frac{p^{2}}{2m}+V-\mu {\cal O})^
{\vphantom{\dagger}}_{j\mu,j^{\prime
}\mu ^{\prime }}\nonumber \\
&&\times(f_{\mu }^{\dagger })^{\bar{a}}(f_{\mu ^{\prime
}})^{b}X_{j^{\prime }}^{\bar{a}}X_{j^{\prime }}^{b}\nonumber \\
&&-\HH_{D}.
\label{mopam:eq}
\end{eqnarray}
Thus, within this approximation, the single-ion energies
$\tilde{E}_{j\Gamma}^{(n)}$ are shifted by the self-consistent 
field of interaction with other
ions, delocalized electrons and the localized ones, but belonging to
other
sites: $\tilde{E}_{j\Gamma }^{(n)}=$ $E_{j\Gamma }^{(n)}+\langle \Gamma
|\Phi (r)|\Gamma \rangle .$ However, the self-consistent field should be
found from the Poisson equation and, therefore, it contains only the
Hartree
part of the interaction between the collective quasi-particles.
Nevertheless,
this problem does not coincide with the single-particle Hartree
approximation, since it contains additional information about the
structure
of the many-electron states $|\Gamma \rangle $ of the ion. Due to the 
non-linearity 
of the problem the approximations for the Green functions (GFs)  
can be used in a different form from the 
standard single-particle problem in the Hartree approximation. This does
not only 
lead to different expressions for the charge density, but makes it
possible
to
improve the approximations for the matrix elements. Thus, in spite of
the
fact that the static field $\Phi (\rr)$ fulfills the Poisson equation, 
in this approximation $V_{jLj^{\prime }L^{\prime }}$ does not
necessarily
correspond to the potential for the electrons in the Hartree
approximation.
Particularly, as seen from the solution of the self-consistent Anderson
model,
presented in Ref.~\cite{III}, the potential depends implicitly on the
many-body population numbers of the ion states already in the lowest 
approximation.

The exchange contribution appears in first order between
electrons via the fluctuation of the field $\varphi $, $\propto
\langle
{\cal T}\delta \varphi (\tau )\delta \varphi (\tau ^{\prime })\rangle $. This
study
we leave for the future, however, it is interesting to note that this
exchange
involves also the contributions from fluctuations caused by the
intra-ion
transitions.

The charge density for the Poisson equation can be found from the GFs. 
It is clear that if it is possible to approximate this potential by a 
spherically symmetrical one, $V_{j\mu j\mu ^{\prime }}=\delta _{\mu \mu
^{\prime }}V_{j},$ then $(f_{\mu }^{\dagger }f_{\mu })^{\xi }=\delta
_{\xi
,[\Gamma ,\Gamma ]}$ for the orbitals, $\mu$, occupied in the state 
$\Gamma $. The averages $\langle c^{\dagger }c\rangle $, $\langle
c^{\dagger
}X\rangle $, $\langle X^{\dagger }c\rangle $, $\langle X^{\dagger
}X\rangle $
needed for finding $\langle \rho (\rr)\rangle $, can be found from the 
corresponding GFs.

\section{Mixing Interaction}
Let us now rewrite the problem in a form close to the standard
periodical Anderson model. This allows us to discuss different possible
definitions for the matrix element of the mixing interaction. Usually
the
Hamiltonian of the Anderson model is written in the following form 
\begin{eqnarray}
&&\hspace*{-5mm}
\HH^{(And)}=\sum_{\kk\sigma }\varepsilon _{\kk}^{\sigma }c_{\kk\sigma
}^{\dagger
}c^{\vphantom{\dagger}}_{\kk\sigma }+\sum_{\kk\sigma }[V_{\kk\mu
}^{\sigma }
e^{i\kk\cdot\RR_{j}}c_{\kk\sigma
}^{\dagger }f^{\vphantom{\dagger}}_{j\mu }+H.c.]  \nonumber \\
&&+\sum_{j\{\mu \}}U_{\mu _{1}\mu _{2}\mu _{3}\mu _{4}}f_{j\mu
_{1}}^{\dagger }f_{j\mu _{2}}^{\dagger }f^{\vphantom{\dagger}}_{j\mu
_{3}} f^{\vphantom{\dagger}}_{j\mu _{4}}
+\sum_{j\mu}\epsilon_{\mu}f_{j\mu}^{\dagger}
f^{\vphantom{\dagger}}_{j\mu}.
\end{eqnarray}
Here $\varepsilon_{\kk}^{\sigma}$ is the spectrum of the conduction
electrons, $V_{\kk\mu}^{\sigma}$ is the matrix element of mixing
interaction, 
\begin{equation}
V_{\kk\mu }^{\sigma}=\int
\dd\rr\phi_{\kk}^{\ast\sigma}(\rr)[\frac{p^{2}}{2m}%
+V(\rr)]\varphi _{\mu }(\rr-\RR_{j}),
\end{equation}
and $U_{\mu _{1}\mu _{2}\mu _{3}\mu _{4}}$ is the matrix elements of
the
Coulomb intra-shell interactions. For the mixing no problem arises in
the case
of the 
impurity Anderson model, since the potential in this case is the
difference
between the periodical potential for the conduction electrons and the
potential
of the impurity, although the local on-site term is equal to zero due to
symmetry
reasons, overlap with the orbitals of neighboring ions gives non-zero
contribution (see the paper by Anderson~\cite{anderson_imp}). In the
case of the periodic Anderson model (PAM) this form of the matrix
element
suggests that either the potential has different symmetry from the Bloch
wave functions, or the functions $\phi_{\kk}^{\ast \sigma }(r)$ and  
$\varphi_{\mu}(\rr-\RR_{j})$ are not orthogonal. It is not clear how to
fulfill the 
first assumption in the case of elemental metals (like Ce metal) since
we
are dealing with a periodic system. In the second case 
\begin{eqnarray}
&&\hspace*{-5mm}
\int\dd\rr\phi_{\kk}^{\ast\sigma}(\rr)[\frac{p^{2}}{2m}+V(\rr)]
\varphi^{\vphantom{\dagger}}_{\mu
}(\rr-\RR_{j})\nonumber \\
&&=\varepsilon_{\kk}^{\sigma}\int
\dd\rr\phi_{\kk}^{\ast\sigma
}(\rr)\varphi^{\vphantom{\dagger}}_{\mu}(\rr-\RR_{j})
=\varepsilon_{\kk}^{\sigma}
{\cal O}^{\vphantom{\dagger}}_{\kk\sigma ,j\mu}.
\end{eqnarray}
Therefore, in order to have a non-zero mixing, one has to work in terms
of an 
non-orthogonal basis set, but in this case a) there is a contribution 
from the chemical potential to the partition function which affects the
mixing, and  
b) the non-orthogonality causes non-zero
anticommutation
relations between the $f$- and $c$-operators. This is never taken into
account in model calculations. 

Let us consider the consequences from our formulation. Since  the mixing
matrix
element has a single-electron form, let us make in our
saddle-point Hamiltonian a transformation
that diagonalizes the conduction
electrons. First we have to rewrite the Hamiltonian,
Eq.(\ref{mopam:eq}) 
in $\kk$-space (for the formulas to be transparent,  
we will write them for the case of one atom in the 
elementary cell, then, the single-ion matrix elements and energies do
not
depend on the ion index). Using the definition:
\begin{equation}
c^{\vphantom{\dagger}}_{jL}=\sum_{\kk}e^{-i\kk\cdot\RR_{j}}
c^{\vphantom{\dagger}}_{\kk L},\;c_{jL}^{\dagger
}=\sum_{k}e^{i\kk\cdot\RR_{j}}c_{\kk L}^{\dagger },
\end{equation}
we find
\begin{eqnarray}
&&\hspace*{-2mm}\tilde{\HH} =\int \dd\qq\;q^{2}\Phi(\qq)\Phi(-\qq)
\nonumber \\
&&+\sum
[h_{L,L^{\prime
}}(\kk)-\mu \,{\cal O}_{L,L^{\prime }}(\kk)]c_{\kk L}^{\dagger }
c^{\vphantom{\dagger}}_{\kk L^{\prime }} \nonumber \\
&&+\sum (\tilde{E}_{\Gamma }^{(n)}-n\mu )h_{j}^{\Gamma }+\sum [V_{\mu
,\mu ^{\prime }}(f_{\mu }^{\dagger }
f^{\vphantom{\dagger}}_{\mu ^{\prime }})^{\xi \neq \lbrack
\Gamma
,\Gamma ]}Z_{j}^{\xi }+h.c.]  \nonumber \\
&&+\sum (\frac{p^{2}}{2m}+V-\mu {\cal O})^{\vphantom{\dagger}}_
{jL,j^{\prime }\mu ^{\prime
}}e^{i\kk\cdot\RR_{j}}(f_{\mu ^{\prime }})^{a}c_{\kk L}^{\dagger
}X_{j^{\prime
}}^{a} 
\nonumber \\
&&+\sum (\frac{p^{2}}{2m}+V-\mu {\cal O})^{\vphantom{\dagger}}_
{j\mu ,j^{\prime }L^{\prime
}}e^{-i\kk\cdot\RR_{j}}(f_{\mu }^{\dagger })^{\bar{a}}X_{j}^{\bar{a}}
c^{\vphantom{\dagger}}_{\kk L}] \nonumber \\
&&+\sum (1-\delta _{jj^{\prime }})(\frac{p^{2}}{2m}+V-\mu {\cal O})^
{\vphantom{\dagger}}_{j\mu,j^{\prime
}\mu ^{\prime }}(f_{\mu }^{\dagger })^{\bar{a}}
(f^{\vphantom{\dagger}}_{\mu ^{\prime
}})^{b}X_{j^{\prime }}^{\bar{a}}X_{j^{\prime }}^{b}\nonumber \\
&&-\HH_{D}.
\end{eqnarray}
Now we have to rewrite this in terms of the operators 
$\alpha^{\vphantom{\dagger}} _{\kk\gamma
}=Z_{\kk}^{\gamma L}c^{\vphantom{\dagger}}_{\kk L},\;
\alpha_{\kk \gamma }^{\dagger }=
c_{\kk L}^{\dagger}\bar{Z}_{\kk}^{L\gamma }$. 
It is easy to see that they are orthogonal to each other 
\begin{eqnarray}
&&\alpha^{\vphantom{\dagger}} _{\kk\gamma }
\alpha _{\kk\gamma ^{\prime }}^{\dagger }+\alpha
_{\kk\gamma
^{\prime }}^{\dagger }\alpha^{\vphantom{\dagger}}_{\kk\gamma
}=Z_{\kk}^{\gamma
L}(c^{\vphantom{\dagger}}_{\kk L}c_{\kk L^{\prime }}^{\dagger }+
c_{\kk L^{\prime }}^{\dagger
}c^{\vphantom{\dagger}}_{\kk L})\bar{Z}_{\kk}^{L^{\prime }\gamma
^{\prime }}  
\nonumber \\
&=&Z_{\kk}^{\gamma L}({\cal O}^{-1})^{\vphantom{\dagger}}_{LL^{\prime }}
\bar{Z}_{\kk}^{L^{\prime }\gamma
^{\prime}}=Z_{\kk}^{\gamma L}(Z_{\kk}^{-1})^{L\gamma _{1}}(\bar{Z}%
^{-1})^{\gamma_{1}L^{\prime }}\bar{Z}_{\kk}^{L^{\prime }\gamma ^{\prime
}}\nonumber \\
&&\hspace*{2cm}=\delta_{\gamma\gamma^{\prime}}.
\end{eqnarray}
\begin{eqnarray}
\mathbf{c}^{\dagger }\mathbf{(h-\mu \,O)c} &=&\mathbf{c}^{\dagger
}[\mathbf{%
\bar{Z}(\bar{Z}}^{-1}\mathbf{hZ}^{-1})\mathbf{Z-}\mu
\mathbf{\,\bar{Z}Z]c=} 
\nonumber \\
(\mathbf{c}^{\dagger }\mathbf{\bar{Z})[\tilde{h}-}\mu \mathbf{I](Zc)}
&=&%
\mathbf{\alpha }^{\dagger }\mathbf{[\tilde{h}-}\mu \mathbf{I]\alpha }.
\end{eqnarray}
Thus, we get
\begin{eqnarray}
&&\hspace*{-5mm}
\tilde{\HH} =\int \dd\qq\;q^{2}\Phi (\qq)\Phi (-\qq)\nonumber \\
&&+\sum [(\bar{Z}%
_{\kk}^{-1})^{\gamma L}h_{L,L^{\prime }}(\kk)(Z_{\kk}^{-1})^{L\gamma
^{\prime
}}-\mu \,\delta_{\gamma \gamma ^{\prime }}]\alpha _{\kk\gamma }^{\dagger
}\alpha^{\vphantom{\dagger}} _{\kk\gamma ^{\prime }}  \nonumber \\
&&+\sum (\tilde{E}_{\Gamma }^{(n)}-n\mu )h_{j}^{\Gamma }+\sum [V_{\mu
,\mu
^{\prime }}(f_{\mu }^{\dagger }f_{\mu ^{\prime }})^{\xi \neq \lbrack
\Gamma
,\Gamma ]}Z_{j}^{\xi }+h.c.]  \nonumber \\
&&+\sum (\bar{Z}^{-1})^{\gamma L}(\frac{p^{2}}{2m}+V-\mu {\cal O})^
{\vphantom{\dagger}}_{jL,j^{\prime
}\mu ^{\prime }}e^{i\kk\cdot\RR_{j}}
(f^{\vphantom{\dagger}}_{\mu ^{\prime }})^{a}\alpha _{\kk\gamma
}^{\dagger }X_{j^{\prime }}^{a}  \nonumber \\
&&+\sum (\frac{p^{2}}{2m}+V-\mu {\cal O})^{\vphantom{\dagger}}_{j\mu ,j^{\prime
}L}(Z_{\kk}^{-1})^{L\gamma
}(f_{\mu }^{\dagger })^{\bar{a}}e^{-i\kk\cdot\RR_{j}}X_{j}^{\bar{a}}
\alpha^{\vphantom{\dagger}}_{\kk\gamma }
\nonumber \\
&&+\sum (1-\delta _{jj^{\prime }})(\frac{p^{2}}{2m}+V-\mu {\cal O})^
{\vphantom{\dagger}}_{j\mu,j^{\prime
}\mu ^{\prime }}(f_{\mu }^{\dagger })^{\bar{a}}
(f^{\vphantom{\dagger}}_{\mu ^{\prime
}})^{b}X_{j^{\prime }}^{\bar{a}}X_{j^{\prime }}^{b}\nonumber \\
&&-\HH_{D}.
\end{eqnarray}
At last, diagonalizing the conduction electron Hamiltonian
$\mathbf{\bar{Z}}^{-1}\mathbf{hZ}^{-1}$ , we have 
\begin{eqnarray}
&&\hspace*{-2mm}
\tilde{\HH}=\int \dd\qq\;q^{2}\Phi (\qq)\Phi (-\qq)+\sum [\varepsilon
_{\kk\lambda }-\mu \,]\tilde{c}_{\kk\lambda }^{\dagger }
\tilde{c}^{\vphantom{\dagger}}_{\kk\lambda } \nonumber \\
&&+\sum (\tilde{E}_{\Gamma }^{(n)}-n\mu )h_{j}^{\Gamma }+\sum [V_{\mu
,\mu
^{\prime }}(f_{\mu }^{\dagger }
f^{\vphantom{\dagger}}_{\mu ^{\prime }})^{\xi \neq \lbrack \Gamma
,\Gamma ]}Z_{j}^{\xi }+h.c.]  \nonumber \\
&&+\sum V_{\lambda
}^{\bar{a}}(k)e^{i\kk\cdot\RR_{j}}\tilde{c}_{\kk\lambda
}^{\dagger
}X_{j^{\prime }}^{a}  \nonumber \\
&&+\sum e^{-i\kk\cdot\RR_{j}}X_{j}^{\bar{a}}
\alpha^{\vphantom{\dagger}} _{\kk\gamma} \nonumber \\
&&+\sum (1-\delta _{jj^{\prime }})(\frac{p^{2}}{2m}+V-\mu {\cal O})^
{\vphantom{\dagger}}_{j\mu
,j^{\prime }\mu ^{\prime }}(f_{\mu }^{\dagger })^{\bar{a}}
(f^{\vphantom{\dagger}}_{\mu^{\prime
}})^{b}X_{j}^{\bar{a}}X_{j^{\prime}}^{b}\nonumber \\
&&-\HH_{D},
\end{eqnarray}
where the band energy and mixing matrix elements are   
\begin{equation}
\varepsilon_{\kk\lambda }=\vartheta _{\gamma }^{\ast \lambda
}(\kk)(\bar{Z}
_{\kk}^{-1})^{\gamma L}
h^{\vphantom{\dagger}}_{L,L^{\prime }}(\kk)(Z_{\kk}^{-1})^{L\gamma
^{\prime
}}\vartheta _{\gamma ^{\prime }}^{\lambda }(\kk),
\end{equation}
\begin{eqnarray}
V_{\lambda }^{\bar{a}}(\kk) &=&\vartheta _{\gamma }^{\ast \lambda
}(\kk)(\bar{Z}
^{-1})^{\gamma L}(\frac{p^{2}}{2m}+V-\mu {\cal O})^
{\vphantom{\dagger}}_{jL,j^{\prime }\mu ^{\prime
}}(f^{\vphantom{\dagger}}_{\mu ^{\prime }})^{a}\nonumber \\
&\equiv& v_{\mu ^{\prime 
}}^{\lambda }(\kk)(f^{\vphantom{\dagger}}_{\mu
^{\prime }})^{a}, \\
V_{\lambda }^{\ast \bar{a}}(\kk) &=&(\frac{p^{2}}{2m}+V-\mu {\cal O})^
{\vphantom{\dagger}}_{j\mu
,j^{\prime }L}(Z_{\kk}^{-1})^{L\gamma }\theta _{\gamma }^{\lambda
}(\kk)(f_{\mu }^{\dagger })^{\bar{a}}\nonumber \\
&\equiv& (f_{\mu }^{\dagger
})^{\bar{a}
}v_{\mu ^{\prime }}^{\ast \lambda }(\kk),
\end{eqnarray}
where $\{\vartheta\}$ diagonalizes 
$\mathbf{\bar{Z}}^{-1}\mathbf{hZ}^{-1}$, 
and $v_i^{\lambda}$ is the one-electron hybridization parameter
Thus, the matrix element of the mixing interaction has to be found
self-consistently together with the charge density (which in turn
depends on
the
particular approximation in which the PAM is solved) and can be
represented
in the form of a sum over all localized orbitals of products of the
matrix
element of the single-particle potential (on the conduction electron
Bloch
function $\phi _{k\lambda }(\rr)$  and the localized orbital $\chi _{\mu
}(\rr)$) and the factor $(f_{\mu ^{\prime }})^{a}$, which reflects the
contribution of this orbital into Fermion-like transition. 
Irkhin~\cite{irkhiny} has performed the calculation of this factor
 for the $4f$-elements  making use of  the Racah technique leaving,
the single-particle matrix element undefined. 
Putting $\bar{a}=\bar{a}(\Gamma_n,\Gamma_{n-1})$ and $\mu =
(l,m,\sigma)$
and using the result of the calculation of Irkhin, we can write these
coefficients as follows
\begin{eqnarray}
&&\hspace*{-5mm}
(f^{\dagger}_{\mu })^{\bar{a}} =
\langle \Gamma_n |f^{\dagger}_{lm\sigma }|\Gamma_{n-1}\rangle\nonumber
\\
&&=\sqrt{n[\Gamma_n][\Gamma_{n+1}]}\left \{
  \begin{array}{ccc}
  S_n     & L_n     & J_n     \\
  S_{n-1} & L_{n-1} & J_{n-1} \\
  1/2     & l       & j
  \end{array}
\right \} G^{\Gamma_n}_{\Gamma_{n-1}},
\end{eqnarray}
where $\gamma=(lm_l\sigma)$ are the one-electron quantum numbers,
$[a]=2a+1$ and $G^{\Gamma_n}_{\Gamma_{n-1}}$ is the parentage
Racah coefficients and $j=l+1/2$. The coefficients
$(f^{\dagger}_{\mu })^{\bar{a}}$ are different
for the two channels, $j=5/2$ and $j=7/2$.
The square of them are given in
Table.~\ref{irkhin_coeffs:tab}.
\begin{table}
\begin{tabular}{ccccccccccccccc}
n & 1 & 2 & 3 & 4 & 5 & 6 & 7 & 8 & 9 & 10 & 11 & 12 & 13 & 14 \\ \hline
$[f_{5/2}]^2$ &
1 & 2.25 & 2.3 & 1.8 & 0.9  & 0.15 & 0   & 1.6 & 2.3 & 4.1 & 1.7 &
0.9 & 0.3 & 0 \\
$[f_{7/2}]^2$ &
0 & 0.25 & 0.6 & 0.8 & 0.63 & 0    & 0.9 & 0.5 & 1.3 & 4.1 & 3.0 &
3.3 & 2.8 & 1 \\
\end{tabular}
\caption{Coefficients for the mixing interaction for many-body states.
For the $f$-series, taken from~\cite{irkhiny}.}
\label{irkhin_coeffs:tab}
\end{table}
However, there are more considerations to the problem, 
since the Bloch orbitals are not orthogonal to
the localized orbitals. In the equation of motion for the operators
appear
combinations involving the overlap matrices. For the Hamiltonian which
includes hopping and mixing interactions (the periodic Hubbard-Anderson
model)
the diagram technique and different approximations for the Green 
functions (GF) 
are given in I, while the full self-consistent solution for rare earths
in 
the simplest possible approximation is presented in Ref.~\cite{III}. 

\section{Discussion}
Many different suggestions exists in the literature on how the 
parameters of the Anderson Hamiltonian (periodical or impurity)  
should be calculated from an {\em ab initio} approach  
and a consensus is not yet achieved. This motivated
us to make an attempt to derive the parameters from a total Hamiltonian.
Since the Anderson model is usually used for the description of the
cases when
strong electron correlations are well developed, it is reasonable to
consider strong intra-atomic Coulomb interactions first. For this reason
we have performed a derivation of the self-consistent PAM in four
steps:
i) first we separate, in the Hamiltonian, the strong intra-atomic
interactions and approximately diagonalized them with the help of
many-electron functions describing different ion terms; ii) we
expressed all operators, describing $f$- and core
electrons in terms of Hubbard operators; iii) we performed
the Hubbard-Stratanovich decoupling of the Coulomb interaction and
finally
iv) we found the equation for the electric potential in a saddle-point
approximation.
This lead us to the effective Hamiltonian, which coincide with the
generalized PAM.

All three operations before making the saddle-point approximation are
exact. However, the saddle-point approximation neglects 
the contributions from the exchange interaction for the delocalized
particles
(excitations) and Coulomb screening effects, which appear only in the
next 
orders with respect to fluctuations of this field near its saddle-point
value. Thus, we may conclude that from this point of view 
this model is quite rough. On the other hand, the way suggested here 
has an obvious advantage compared to the PAM with non-self-consistent 
parameters since it, at least, takes care about perturbations 
of the local charge density 
which may introduce quite large changes in energy. 

\section*{Acknowledgment}
We are grateful to The Natural Science Foundation (NFR
and TFR), The Swedish Foundation
for Strategic Research (SSF) and the G\"oran Gustafsson foundation for
support.

\end{document}